# Highly strained Ge micro-blocks bonded on Si platform for Mid-Infrared photonic applications

A. Gassenq[1], K.Guilloy[1], N. Pauc[1], D. Rouchon[2], J. Widiez[2], J. Rothman[2], J-M. Hartmann[2], A. Chelnokov[2], V. Reboud[2], V. Calvo[1]

1 - CEA-INAC, Univ. Grenoble Alpes, 38000 Grenoble France

2 - CEA-LETI, Univ. Grenoble Alpes, 38000 Grenoble, France

Adding sufficient tensile strain to Ge can turn the material to a direct bandgap group IV semiconductor emitting in the mid-infrared wavelength range. However, highly strained-Ge cannot be directly grown on Si due to its large lattice mismatch. In this work, we have developed a process based on Ge micro-bridge strain redistribution intentionally landed to the Si substrate. Traction arms can be then partially etched to keep only localized strained-Ge micro-blocks. Large tunable uniaxial stresses up to 4.2% strain were demonstrated bonded on Si. Our approach allows to envision integrated strained-Ge on Si platform for mid-infrared integrated optics.

Silicon photonics merges optical and electronic components to be integrated together onto a single microchip. Since the great interest in Si photonics has been demonstrated for telecommunication application more applications have emerged in the Mid-Infrared (MIR) wavelength range (2 to 5µm) like gas sensing [1]. Since the $SiO_2$ is transparent up to 3.5 µm wavelength [2], Silicon-On-Insulator (SOI) MIR spectrometers have been successfully demonstrated [3]. For longer wavelength, Ge on Si platform is instead used [4], [5]. However, the monolithically integration of active devices (sources and photodetector) are still missing to unlock all the possibilities for Si photonics in the MIR. The current solution consist of growing [6], [7] or bonding III-V materials [8], [9] on Si but these heterogeneous integration are not considered as Complementary Metal Oxide Semiconductor (CMOS) compatible by industrial foundries. For the group IV material sources, doped-Ge [10] and GeSn alloys [11] are envisioned to tackle the challenge to obtain a laser source CMOS compatible. Note that GeSn devices are also studied for photo-detection applications on Si substrate [12], [13]. Strained-Ge structures exhibits also a MIR direct bandgap [14], [15] but their integration, which require extra process step to localize and contact the strained-Ge, remains challenging. Such integration on a Si platform would open the way to MIR fully integrated active devices (source and detectors) which are CMOS compatible. The availability of such approach would have many applications for sensing systems [2] or on-chip optical interconnects for high-performance computing [16].

In this work, we propose to bond highly strained Ge directly on Si using a specific processing based on micro-bridge landing followed by the traction arm etching. Compared to free standing membranes, our landing approach will greatly facilitate the next process steps needed to go towards electrically pumped laser sources or integrated photo-detectors based on strained-Ge for MIR applications. We measure a Raman spectral shift of around 8 cm$^{-1}$ corresponding to 4.2% in a Ge micro-block directly bonded on Si which is higher than the first demonstration of 3 cm$^{-1}$ using the biaxial approach [17].

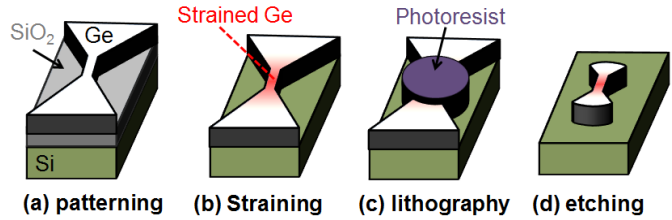

**Figure 1**: *Processing of a) Ge layer patterning; b) SiO$_2$ under-etching leading to concentrate the strain in the centre of landed micro-bridge; c) final lithography and d) to remove traction arms*

The Figure 1 presents the device process flow. In order to amplify strain in the germanium, we have used the strain re-distribution principle [18], [19] performed in our 200 mm GeOI wafer [20], [21]. The Ge layers are 0.35 µm thick on 1 µm SiO$_2$ on a Si substrate. Bridge patterning was performed using e-beam lithography followed by dry etching in an Inductively Coupled Plasma (ICP) reactor with Cl$_2$, N$_2$ and O$_2$ gas (Figure 1-a). After the Ge etching, the bridge is landed to the Si substrate [22] using a low speed under-etching recipe at 50°C in a dedicated HF vapour reactor (Figure1-b). The traction arms are then partially removed thanks to a last lithography step (Figure 1-c) followed by the etching of the Ge with the same ICP etching recipe (Figure 1-d). The design parameters of the device are presented in the Figure 2-a. The Figure 2-b presents a Scanning Electron Microscopy imaging of a fabricated device at the end of the processing when the Ge micro-block is directly bonded on the Si.

Figure 2-c presents a Raman spectral shift mapping measured by micro-Raman spectroscopy on a Ge micro-block (e=4 µm, L=300 µm, d=12 µm and C=100 µm in the Figure 2-b). The Raman spectral shift were measured with an input laser at 785 nm wavelength (given a measured depth of 200 nm [23]) with a diameter spot of around 1 µm. The used laser intensity was low enough to avoid heating effect during the measurements [24]. Each Raman shift has been fitted using a Lorentz function. A bulk Ge substrate was used as reference to measure 0 % strain to quantify the Raman spectral shifts which increase with strain [25]. We observe here that the strain is concentrated in the central region while the traction arms only undergo a much smaller strain, as observed for free standing membranes [18], [26].

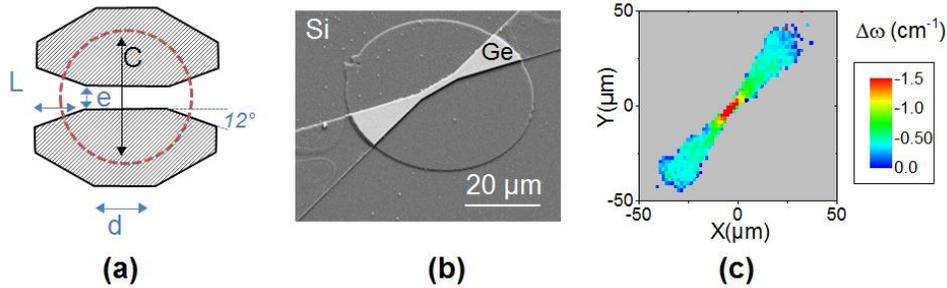

**Figure 2**: *(a) Design parameter used to tune the strain in the Ge micro-bridge; (b) Fabricated Ge micro-block landed/bonded on Si; (c) Raman spectroscopy mapping of a Ge micro-block bonded on Si, the strain is still concentrated in the narrowest part of the micro-block*

Figure 3 presents the Raman spectra before and after the arm removal for two representative designs: when the traction arms are completely removed (C<d in the Figure 2-a) or when the traction arms are partially etched (C>d in the Figure 2-a). Figure 3a-c presents a micro-bridge with the following dimension e=1 µm, L=110 µm, d=12 µm and C=10 µm. After the under-etching, the measured Raman spectra shift has been evaluated at 5 cm$^{-1}$ compared to the Ge (Figure 3 a-b). The traction arms have been then fully removed which results in a strain relaxation leading to a Raman spectrum close to the one for bulk Ge wavenumber (Figure 3-c). Figure 3 d-f presents a micro-bridge with the following dimension e=0.5 µm, L=130 µm, d=8 µm and C=30 µm. After under-etching, the measured Raman spectra shift is evaluated to be 8 cm$^{-1}$ (Figure 3-d-e). Then, we partially removed the traction arms. In that case, the strain is conserved (Figure 3-f) thanks to the bonding between the partial arms and the substrate, which is probably due to the small thermal annealing (50°C) which occurs during the HF vapour under-etching. Therefore we obtain a Ge structure which is bonded to the Si, with a very high strain of 4.2% (corresponding to a Raman shift of 8 cm$^{-1}$ [23]). This value is very close to the actual maximum reported Ge strain of 4.9% [23] and higher than the previous demonstration of 0.7% (3 cm$^{-1}$) using the biaxial bonding approach [17].

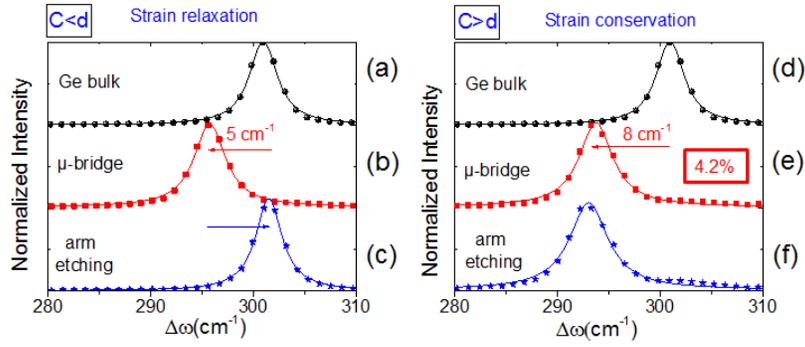

**Figure 3:** *Normalized Raman spectra of Ge bulk, Ge micro-bridge and Ge micro-block with (a,b,c) the traction arms completely etched; and with (d,e,f) the traction arms partially etched.*

Photoluminescence (PL) measurements were also performed on strained-Ge micro-blocks. A 1047 nm Nd:YLF Q-switched laser, outputting an average power of 1 mW on an area 20 µm wide, was used to pump the material. The luminescence was collected through a Cassegrain objective and analysed using a home-built FTIR spectrometer. A CdHgTe avalanche photodiode was used to detect the emitted light. Figure 4 shows the photoluminescence spectra of four Ge micro-blocks: a relaxed (Figure 4-a) and three differently strained Ge micro-blocks (Figure 4 b-d). Note that the PL measurements were not performed on the highest strained Ge micro-blocks presenting a Raman shift of 8 cm$^{-1}$ (Fig3-e) as the dimension of the strained region are too small to obtain a correct PL signal with the sensitivity of our detection system. For strained samples two peaks are detected corresponding to the transition between the conduction band at Γ and the light- and heavy-hole valence bands [27]. The strain is tuned here by the design of the micro-bridges [23] up to 3.4 % with a peak emission energy red-shifting when strain is increased, as expected [18]. However, for strain relaxed samples (Figure 4-a), the emission peak is shifted back at 0.80 eV, corresponding to the energy of the direct band gap of unstrained Ge.

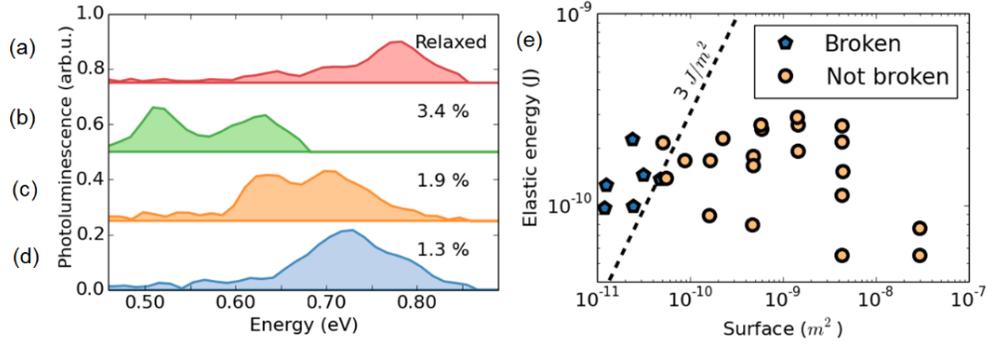

**Figure 4:** *Photoluminescence spectra of integrated strained-Ge micro-block for a) a relaxed sample and at b) 3.4 %, c) 1.9 % and d) 3.4 %; e) total surface of the micro-bridges plotted as a function of the total elastic energy of the stressed material. Samples marked in light yellow circle are strained while samples marked in dark blue hexagon released their strain.*

Figure 4-e shows the elastic energy plotted as a function of the total surface (traction arms and strained area) of all devices. The elastic energy of their materials is written as:

$$E = \tfrac{1}{2}\, \sigma\, \varepsilon\, V$$

where $\sigma$ is the stress, $\varepsilon$ the strain and V the volume of the central region. We observe that the samples relax when the bonding surface is too small compared to the strain (dark blue hexagon of figure 4-e). The samples are expected to relax when their elastic energy, divided by the area in contact with the substrate, exceeds a certain threshold. From our set of samples, we estimate this elastic energy threshold to be at around 3 J/m² which is in the same order of magnitude compared to the bonding energies measured from wafer bonding experiments [28]–[30]. We hypothesize that further thermal annealing could increase this elastic energy threshold.

In this article, we demonstrate the bonding of highly strained Ge structures to a Si substrate. A maximum of 4.2% induced strain is reported, which is reached by keeping partially the traction arm of landed Ge micro-bridges. This work allows envisioning compact strained-Ge integration on Si for photonics applications.

*Acknowledgements:* The authors would like to thanks the Plateforme de Technologie Amont and 41 in Grenoble for the clean room facilities. This work was supported by the CEA Phare project photonics.